\documentclass[aps,pra,superscriptaddress,twocolumn,eqsecnum,showpacs]{revtex4}
\usepackage{amsmath,bm}
\usepackage{graphicx,epsfig,braket,amssymb,amsfonts,color}

\newcommand{\ba}{\begin{eqnarray}}
\newcommand{\ea}{\end{eqnarray}}
\newcommand{\be}{\begin{equation}}
\newcommand{\ee}{\end{equation}}
\newcommand{\bd}{\begin{displaymath}}
\newcommand{\ed}{\end{displaymath}}
\renewcommand{\v}[1]{{\bf #1}}
\newcommand{\bpm}{\begin{pmatrix}}
\newcommand{\epm}{\end{pmatrix}}
\newcommand{\nn}{\nonumber \\}

\begin{document}

\title{Arbitrary Chern number generation in the three-band model from momentum space}
\author{Soo-Yong Lee}
\affiliation{Department of Physics, Sungkyunkwan University, Suwon
440-746, Korea}
\author{Jin-Hong Park}
\affiliation{Center for Emergent Matter Science, RIKEN, Wako, Saitama 351-0198, Japan}
\author{Gyungchoon Go}
\affiliation{Department of Physics, Sungkyunkwan University, Suwon
440-746, Korea}
\author{Jung Hoon Han}
\email[Electronic address:$~~$]{hanjh@skku.edu}
\affiliation{Department of Physics, Sungkyunkwan University, Suwon 440-746, Korea}

\begin{abstract} A simple, general rule for generating a three-band model with arbitrary Chern numbers is given. The rule is based on the idea of monopole charge-changing unitary operations and can be realized by two types of simple unitary operations on the original Hamiltonian. A pair of monopole charges are required to produce desired topological numbers in the three-band model. The set of rules presented here offers a way to produce lattice models of any desired Chern numbers for three-sublattice situations.
\end{abstract}
\maketitle

\section{Introduction}

Topological theory of the band structure began, albeit implicitly, with the work of
Thouless, Kohmoto, den Nijs, and Nightingale on the quantized Hall conductance in
a two-dimensional model of electron motion~\cite{TKNN}. The transverse conductivity of a
filled band was identified with what is known as the first Chern number in mathematics of the same band. It was immediately established afterwards that general non-interacting bands were characterized by Chern numbers~\cite{Simon}. While the original model analyzed by Thouless {\it et al}., namely the Harper's model~\cite{harper}, contained a myriad of sub-bands and Chern numbers at general rational filling factors, Haldane was later able to produce a simple two-band model that contained a non-zero Chern number $\pm 1$~\cite{Haldane}. Three-band generalization in the context of the kagome lattice model was done by Ohgushi, Murakami, and Nagaosa, where the Chern numbers were +1, 0, -1~\cite{OMN}. Later theoretical studies showed how the Chern numbers in their respective models are related to the Skyrmion structure in the Brillouin zone (See Refs. \onlinecite{varma,go-3band} and references therein).

Search for models with higher Chern numbers began in a flurry in recent years employing a diversity of ideas such as extended hopping and the multi-layer extension~\cite{ran,das-sarma,niu}. Without exception, all these models consider extensions of the original Hamiltonian in the {\it real space}. In this paper, we explore the contrasting path to view the Chern-number-changing process in the {\it momentum space}. It is well known, for two-band models, that a given Chern number corresponds to the monopole charge associated with the two-component spinor wave function~\cite{nakahara}. By exploiting this fact, we establish the existence of certain unitary transformations carrying the extra topological numbers in the form of singular gauge potentials. The unitary transformation provides a convenient path for obtaining lattice models with arbitrary Chern numbers by changing the monopole charge.

The idea of topology-changing unitary operation applies to the three-band case as well, but now one manipulates a pair of monopole charges instead of one as in the two-band problem. There is no shortage of condensed matter and cold atom systems that fit within the three-band framework, as recent vigorous investigation of their topological properties have demonstrated (See Ref. ~\cite{jo} and references cited therein). Our scheme can be applied to existing three-band models as a ready recipe for generating their higher-Chern-number derivatives.

\section{Two-band Chern number engineering: review}
\label{sec:two-band}

We begin by summarizing several well-known facts about the two-band Hamiltonian. An arbitrary two-band Hamiltonian in momentum $(\v k)$ space can be written as

\ba {\cal H} (\v k ) = -\v d(\v k) \cdot \bm \sigma , \label{eq:2.1}\ea
involving the Pauli matrix $\bm{\sigma}= (\sigma^x, \sigma^y, \sigma^z )$ and a smooth three-component vector $\v d(\v k)$, without the uniform part $\varepsilon(\v k)\mathbb{I}_{2\times2}$ that has no effect on the band topology. Any other Hamiltonian related to this one by a unitary
transformation $U(\v k)$,
\ba {\cal H}' (\v k) = {\cal U} (\v k) {\cal H} (\v k) {\cal U}^\dag (\v k), \ea
must possess the same band energies $\pm |\v d(\v k)|$. Wave functions of the new Hamiltonian, $|\psi'_i (\v k)\rangle$, are related to $|\psi_i (\v k)\rangle$ of the old ${\cal H}(\v k)$ through

\ba |\psi'_i  (\v k) \rangle = {\cal U} (\v k) |\psi_i (\v k) \rangle, \ea
where $i$ stands for the band index. The geometric connection~\cite{nakahara,Simon} for the new state is obtained by
\ba \v a'_{i} (\v k) \!=\! \v a_{i} (\v k)  \! -\! i \langle \psi_i (\v k)| {\cal U}^\dagger
(\v k) \partial_{\v k}  {\cal U} (\v k) |\psi_i (\v k) \rangle , \label{eq:vector_potential}\ea
$\v a_{i} (\v k) = -i \langle \psi_i (\v k)|\partial_{\v k}|\psi_i (\v k) \rangle$. The Chern number for the $i$-th band of the new Hamiltonian is therefore to be obtained from integrating the curl of the new gauge potential $\v a'_i (\v k)$ which may, in principle, differ from the old one due to the singular contribution of the unitary matrix.

Writing the two-band model (\ref{eq:2.1}) in the form (dropping the $\v k$ label)
\ba {\cal H} = - d \left(
                       \begin{array}{cc}
                       \cos \theta & e^{-i \phi} \sin \theta\\

                        e^{i \phi} \sin \theta & -\cos \theta  \\
                       \end{array} \right)\label{eq:twoband_Hamiltonian}\ea
where $d$ stands for the magnitude $|\v d|$ and the unit orientation vector $\hat{d} =
(\sin \theta \cos \phi , \sin \theta \sin \phi , \cos \theta )$ is introduced, the two
eigenstates of the Hamiltonian are
\ba |\psi_\uparrow \rangle = \left(
                       \begin{array}{c}
                          \cos \frac{\theta}{2} \\
                        e^{i \phi} \sin \frac{\theta}{2} \\
                       \end{array} \right), ~
|\psi_\downarrow\rangle = \left(\begin{array}{c} -e^{-i \phi} \sin \frac{\theta}{2} \\
                        \cos \frac{\theta}{2} \\ \end{array}
                     \right). \label{eq:cp1}
\ea

\begin{figure}[t]
\includegraphics[width=80mm]{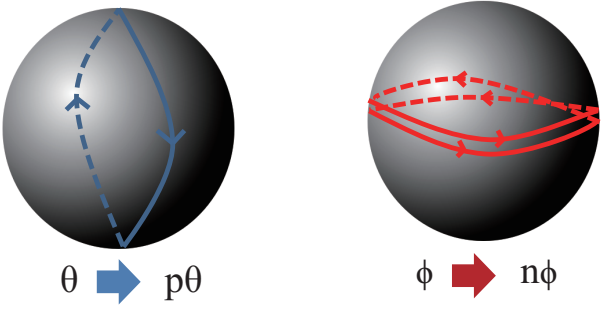}
\caption{(color online) Schematic representation of the monopole charge-changing unitary operation in the two-band model. $U$ operation changes the winding number of the polar angle $\theta$ by $\theta \rightarrow p \theta$ ($p=2$ in the above left). $V$ operation changes the winding number of the azimuthal angle $\phi \rightarrow n \phi$ which increases the Chern number by $n$ ($n=2$ in the above right).}
\label{fig1}
\end{figure}

We now introduce unitary matrices to manipulate the Chern number. The scheme is based on the pedagogical observation that the Hamiltonian ${\cal H}$ is diagonalized with the unitary matrix

\ba U=\bpm \cos {\theta\over 2} & - e^{-i\phi} \sin {\theta\over 2} \\
e^{i\phi} \sin {\theta \over 2} & \cos {\theta \over 2} \epm, ~~ U^\dag {\cal H} U = -d \sigma^z . \label{eq:2.7} \ea
The diagonal Hamiltonian $-d \sigma^z$ carries trivial eigenstates $(1,0)$ and $(0,1)$ and have no topology associated with it. Therefore, the topological charge of ${\cal H}$ must have been planted there by the unitary matrix $U$. In other words, $U$ is a singular gauge that contributes non-zero flux in Eq. (\ref{eq:vector_potential}). And if one such unitary rotation could turn a non-topological Hamiltonian $-d \sigma^z$ into a topological one, what would be the consequence of implementing the same unitary rotation, say $n$ times?

In order to make the discussion fully general we introduce a second unitary rotation that changes the azimuthal angle $\phi$ in the Hamiltonian by one unit, $e^{i\phi} \rightarrow e^{2i\phi}$:

\ba V = \bpm e^{-i \phi/2}  & 0 \\ 0 & e^{i \phi/2} \epm. \label{eq:2.14}\ea
We then combine the two unitary transformations as one general operation,

\ba {\cal U}^{(p,n)} = [V]^{n-1} [U]^{p-1} ,\ea
which results in the change of the Hamiltonian

\ba {\cal{H}}^{(p,n)}&=& -d {\cal U}^{(p,n)} \bpm \cos \theta  &  e^{-i \phi}  \sin \theta \\ e^{i \phi}
\sin \theta & -\cos \theta  \epm [{\cal U}^{(p,n)}]^\dag \nn
 ~~ &=&  -d \bpm \cos p\theta  &  e^{-i n\phi}  \sin p\theta \\ e^{i n\phi} \sin p\theta & -\cos
p\theta  \epm .  \ea
In this notation, $\hat{d}=(\sin \theta \cos \phi, \sin \theta \sin \phi, \cos \theta)$ in the original Hamiltonian becomes $\hat{d}^{(1,1)}$. Generally,

\ba && \hat{d}^{(p,n)} = (\sin p \theta \cos n \phi , \sin p\theta \sin n \phi , \cos p\theta), \nn
&& {\cal H}^{(p,n)} = -d  \hat{d}^{(p,n)} \cdot \bm \sigma .  \ea
The ${\cal U}^{(p,n)}$-transformed wave functions are

\ba |{\psi}_\uparrow^{(p,n)}\rangle &=& {\cal U}^{(p,n)} |\psi_\uparrow^{(1,1)}\rangle
=\left(\begin{array}{c}
e^{-i \frac{n \phi}{2}}\cos \frac{p\theta}{2} \\
e^{i \frac{n \phi}{2}} e^{i \phi} \sin \frac{p\theta}{2} \\
\end{array} \right),\\
|{\psi}_\downarrow^{(p,n)}\rangle&=&{\cal U}^{(p,n)} |\psi^{(1,1)}_\downarrow
\rangle
=\left(
\begin{array}{c}
-e^{-i \frac{n \phi}{2}}e^{-i \phi} \sin \frac{p\theta}{2}  \\
e^{i \frac{n \phi}{2}}\cos \frac{p\theta}{2} \\
\end{array} \right)\nonumber. \label{eq:nthVwaveft}\ea
The reference states $|\psi^{(1,1)}_{\uparrow,\downarrow}\rangle$ are the ones given in Eq. (\ref{eq:cp1}). Vector potentials and the magnetic field through the unit sphere~\cite{nakahara} associated with them are

\ba a^{(p,n)}_{\uparrow,\phi} &=& \frac{1-n\cos p\theta}{2\sin \theta}, ~
a^{(p,n)}_{\uparrow,\theta} =0, \nn
{\v B}^{(p,n)}_{\uparrow} &=& n p \frac{ \sin p \theta}{2\sin \theta} \hat{r} = -{\v B}^{(p,n)}_{\downarrow} . \label{eq:nthV_vectorpotential} \ea
Charge of the magnetic monopole is obtained by integrating $\v B^{(p,n)}_\uparrow$ over the unit sphere,

\ba C^{(p,n)}_{\uparrow}= n P = - C^{(p,n)}_{\downarrow}, ~ ~ P = {1- (-1)^p \over 2} .
\label{eq:UV_Chern_number}\ea
Therefore, a series of two-band Hamiltonians ${\cal H}^{(p,n)}$ related by unitary transformations must have identical dispersions and yet possess distinct Chern numbers

\ba C^{(p,n)}=\frac{1}{4\pi} \int_\mathrm{BZ} ~ \hat{d}^{(p,n)} \cdot  \left(
{\partial \hat{d}^{(p,n)} \over \partial k_x} \times {\partial \hat{d}^{(p,n)} \over \partial
k_y } \right) . \nonumber \ea
The Chern number defined here agrees exactly with the magnetic charge given in Eq. (\ref{eq:UV_Chern_number}). Figure \ref{fig1} summarizes the monopole charge-changing operations discussed in this section.

\section{Three-band Chern number engineering}
\label{sec:three-band}

\subsection{Mathematical preliminary}
\label{sec:three-band-mathematical-preliminary}

Chern numbers are derived from the connection, which in turn is derived from the knowledge of the wave function.  Finding eigenfunctions of a given 3$\times$3 Hamiltonian in closed form is exceedingly difficult. On the other hand, it is much easier to do the reverse: express an arbitrary 3$\times$3 Hamiltonian from the knowledge of all three orthogonal eigenfunctions along with their energies. We build heavily upon the work of Byrd and collaborators~\cite{byrd} to show how to do this, and then apply the idea of unitary transformations to generate arbitrary Chern numbers for the bands.

A Hermitian matrix with three eigenvalues $(e_1 , e_2 , e_3 )$ are
found as the SU(3) rotation,

\ba {\cal H} = {\cal U} \bpm e_1 &  0 & 0 \\ 0 & e_2 & 0 \\ 0 & 0 & e_3 \epm {\cal U}^\dag , ~
{\cal U} \in \mathrm{SU(3)} . \label{eq:SU(3)-rotation}\ea
The general SU(3) ${\cal U}$ can be expressed as a series of U(1) and SU(2) rotations,

\ba {\cal U} = {\cal U} (\alpha, \beta, \gamma) e^{i \delta \lambda_5 /2 } {\cal U}
(a, b, c) e^{i d \lambda_8 /2}, \label{eq:SU(3)-Euler} \ea
where ${\cal U} (\alpha, \beta, \gamma)$ is an element of SU(2) given by

\ba {\cal U} (\alpha, \beta, \gamma ) = e^{i \alpha \lambda_3 /2 } e^{i \beta \lambda_2
/2} e^{i \gamma \lambda_3 /2 }. \ea
Ranges of the parameters are~\cite{byrd}

\ba && 0 \le \alpha, \gamma, a, c < 2 \pi, \nn
&&  0 \le \beta, \delta, b \le {\pi},~~ 0 \le d \le 2 \sqrt{3} \pi.
\label{eq:range-of-angles}\ea
Dependence of the parameters in the unitary matrix ${\cal U}$ and the three energies $e_i$ on the two-dimensional momentum $\v k$ of the Brillouin zone will be implicit throughout the discussion.

We display the representation of Gell-Mann matrices used in this paper:

\ba \lambda_1 = \bpm 0 & 1 & 0 \\ 1 & 0 & 0 \\ 0 & 0 & 0 \epm ,  &&  \lambda_2 = \bpm 0 &
-i & 0 \\ i & 0 & 0 \\ 0 & 0 & 0 \epm , \nn
\lambda_3 = \bpm 1 & 0 & 0 \\ 0 & -1 & 0 \\ 0 & 0 & 0 \epm , &&
\lambda_4 = \bpm 0 & 0 & 1 \\ 0 & 0 & 0 \\ 1 & 0 & 0 \epm , \nn
\lambda_5 = \bpm 0 & 0 & -i \\ 0 & 0 & 0 \\ i & 0 & 0 \epm , &&
\lambda_6 = \bpm 0 & 0 & 0 \\ 0 & 0 & 1 \\ 0 & 1 & 0 \epm , \nn
\lambda_7 = \bpm 0 & 0 & 0 \\ 0 & 0 & -i \\ 0 & i & 0 \epm , &&
\lambda_8 = {1\over\sqrt{3}} \bpm 1 & 0 & 0 \\ 0 & 1 & 0 \\ 0 & 0 & -2 \epm . \ea
Acting out the unitary rotation (\ref{eq:SU(3)-rotation}) reveals that, out of the eight
angles that define the SU(3) matrix, the two angles $c$ and $d$ drop out in the final
expression of the Hamiltonian. The space of 3$\times$3 Hermitian matrices with a fixed set of energies $(e_1  , e_2 , e_3 )$ is equivalent to
SU(3)/(U(1)$\times$U(1)).

Recall that the unitary matrix diagonalizing a particular Hamiltonian consists of the eigenstates of the same Hamiltonian. The matrix elements of ${\cal U}$ in the parametrization (\ref{eq:SU(3)-Euler}), with $c=d=0$, become the three eigenstates of the Hamiltonian,

\ba |\psi_1 \rangle &=& \cos \frac{b}{2}~ |u
\rangle- \sin \frac{b}{2}~ |v \rangle , \nn
|\psi_2 \rangle &=& \sin \frac{b}{2}~ |u
\rangle + \cos \frac{b}{2}~ |v \rangle ,\nn
|\psi_3 \rangle  &=&   | w \rangle,
\label{eq:3-eigenstates}\ea
with energies $e_1, e_2, e_3$, respectively. The three orthonormal basis vectors are

\ba |u \rangle &=& e^{i \frac{1}{2}a} \bpm e^{i \frac{1}{2}(\alpha + \gamma)} \cos
\frac{\beta}{2} \cos \frac{\delta}{2} \\ -e^{-i \frac{1}{2}(\alpha-\gamma)} \sin \frac{\beta}{2} \cos \frac{\delta}{2} \\ - \sin \frac{\delta}{2} \epm ,\nn
|v \rangle &=& e^{-i \frac{1}{2} a} \bpm e^{i \frac{1}{2}(\alpha - \gamma)} \sin \frac{\beta}{2}\\
e^{-i\frac{1}{2} (\alpha+\gamma)} \cos \frac{\beta}{2}  \\ 0 \epm, \nn
|w \rangle &=&\bpm e^{i\frac{1}{2} (\alpha + \gamma)} \cos \frac{\beta}{2}
\sin \frac{\delta}{2} \\ -e^{-i\frac{1}{2} (\alpha-\gamma)} \sin \frac{\beta}{2} \sin \frac{\delta}{2} \\
\cos \frac{\delta}{2} \epm. \label{eq:uvw-basis} \ea
An arbitrary $3\times3$ Hamiltonian having $|\psi_i\rangle$ as eigenstates with energies $e_i$ is

\ba {\cal H} &=& \sum_i e_i |\psi_i \rangle \langle \psi_i | \nn
&=& {1\over 2} \Bigl( [e_1 \!+\! e_2 ] + [e_1 \!-\! e_2 ] \cos b \Bigr) |u\rangle \langle u | \nn
&+& {1\over 2} \Bigl( [e_1 \!+\! e_2 ] - [e_1 \!-\! e_2 ] \cos b \Bigr) |v \rangle \langle v | \nn
&+& {1\over 2} (e_2 \!-\! e_1 ) \sin b  \bigl(|u\rangle \langle v| + |v
\rangle \langle u | \bigr) \nn
&+& e_3 |w \rangle \langle w | . \label{eq:4.8}\ea

As we are mainly concerned with the Chern number structure of the three-state Hamiltonian, we proceed to construct a general formula for the curvature of each eigenstate by adopting the projection operator method of Avron, Seiler, and Simon~\cite{Simon}. The
projection operator for each eigenstate $|\psi_i \rangle$ is, $\bm \lambda = (\lambda_1, \cdots, \lambda_8)$,

\ba P_i &=& |\psi_i \rangle \langle \psi_i |  = {1 \over 3}(1+\sqrt{3} \v n_i \cdot \bm
\lambda) , \nn
\v n_i &=& {\sqrt{3} \over 2}\langle \psi_i | \bm \lambda | \psi_i \rangle .
\ea
One can check that $\v n_i$ is an object on the unit seven-sphere, $\v n_i \in S^7$. The Berry curvature for each $|\psi_i\rangle$ is conveniently expressed with $\v n_i$,

\ba {\cal F}_i &=& -i \Bigl(\langle \partial_{x} \psi_i |\partial_{y} \psi_i \rangle \!-\!
\langle \partial_{y} \psi_i |\partial_{x} \psi_i \rangle\Bigr) \nn
&=&  {4 \over 3 \sqrt{3} }\v n_i \cdot \Bigl( {\partial \v n_i \over
\partial x} \times { \partial \v n_i \over
\partial y}\Bigr) , \label{eq:generalized-MH} \ea
where $(x,y)$ spans the coordinates of the base space. The cross product in the above formula is defined with the aid of structure constants of SU(3),

\ba \lambda_a \lambda_b &=& {2 \over 3} \delta_{ab} + i f_{abc} \lambda_c + d_{abc} \lambda_c ,  \nn
(\v u \times \v v )_a &=& f_{abc} u_b v_c. \ea
The inner product $\cdot$ is defined in the usual way of multiplying the matching components. Equation (\ref{eq:generalized-MH}) for the Berry curvature ${\cal F}_i$ in terms of $\v{n}_i \in S^7$ in the three-band Hamiltonian is a new result of our work. The formula derived in Ref.~\cite{galitski} is equivalent to ours, but more complicated in appearance. We note as a further fact of mathematical curiosity that the star product, $(\v u * \v v)_a =\sqrt{3} d_{abc}u_b v_c$, between a pair of $\v n_i$'s obeys the relation

\ba \v n_i * \v n_j =\delta_{ij} \v n_j + |\varepsilon_{ijk}| \v n_k.
\label{eq:constraintS7}\ea
This is an extension of the result found in Ref. \onlinecite{galitski} which only focused on the $i=j$ case.

This concludes the mathematical preliminary for the discussion of topological number-changing unitary operations in the three-band model. Due to the enormous complexity of the algebra involved in the following discussion, we will be forced to consider a slightly restricted space of Hamiltonians by taking $b=0$ in the parametrization of the unitary matrices. Such restriction will not present any barrier in generating arbitrary Chern numbers for the three-band model.

\subsection{A pair of monopoles in the three-band model}
\label{three band chern number changing}

With the $b=0$ restriction in place, the wave functions become simpler, $|\psi_1 \rangle = |u \rangle$, $|\psi_2 \rangle = |v \rangle$, and $|\psi_3 \rangle = |w\rangle$. The connection for each state is

\ba -i \langle u | \partial_\mu | u \rangle &=& \frac{1}{2} \partial_\mu a +  \frac{1}{2}
\cos \beta  \cos^2 \frac{\delta}{2} ~ \partial_\mu \alpha +   \frac{1}{2}\cos^2
\frac{\delta}{2}  ~\partial_\mu \gamma, \nn
-i \langle v | \partial_\mu | v \rangle &=& - \frac{1}{2}\partial_\mu a -\frac{1}{2}\cos
\beta~ \partial_\mu \alpha - \frac{1}{2}\partial_\mu \gamma, \nn
-i \langle w | \partial_\mu | w \rangle &=& \frac{1}{2}\cos \beta  \sin^2
\frac{\delta}{2}~ \partial_\mu \alpha + \frac{1}{2} \sin^2 \frac{\delta}{2}~\partial_\mu
\gamma. \ea
Their Berry curvatures are

\ba {\cal F}_u dxdy &=&  {1\over 4} \sin \beta (d \alpha \wedge d \beta) \!+\! {1\over 4} \sin
\delta (d \gamma \wedge d \delta)  \nn
&& ~~ \!+\! {1 \over 8} \sin [\beta \!+\! \delta]\Bigl(d \alpha \wedge d (\beta \!+\! \delta)\Bigr)  \nn
&& ~~~~ \!+\! {1 \over 8} \sin [\beta \!-\! \delta]\Bigl(d \alpha \wedge d (\beta \!-\! \delta)\Bigr), \nn
{\cal F}_v dxdy & = & -\frac{1}{2} \sin \beta (d\alpha \wedge d \beta), \nn
{\cal F}_w  dxdy &=& {1\over 4}  \sin \beta (d \alpha \wedge d \beta) \!-\! {1\over 4} \sin
\delta (d\gamma \wedge d \delta )  \nn
&& ~~ \!-\! {1 \over 8}\sin [\beta\!+\!\delta] \Bigl(d \alpha \wedge d (\beta \!+\! \delta)\Bigr) \nn
&& ~~~~ \!-\! {1 \over 8}\sin [\beta\!-\!\delta] \Bigl(d \alpha \wedge d (\beta \!- \!\delta)\Bigr) . \label{eq:Fxy-uvw}\ea
A wedge product $\wedge$ is defined by

\ba d \alpha \wedge d \beta \!=\! \Bigl( {\partial \alpha \over \partial x} {\partial
\beta \over \partial y} \!- \!{\partial \alpha \over
\partial y} {\partial \beta \over \partial x} \Bigr) dx dy
\!=\! - d \beta \wedge d \alpha. \ea
The three Berry curvatures add up to zero~\cite{Simon}. Let us introduce four ``integers" (integrals of solid angles over a unit sphere divided by $4\pi$)

\ba N_1 &=& {1 \over 4\pi}\int \sin \beta (d \alpha \wedge d \beta),\nn
N_2 &=& {1 \over 4\pi}\int \sin \delta (d \gamma \wedge d \delta ), \nn
N_3 &=& {1 \over 4\pi}\int \sin [\beta+\delta] \Bigl(d \alpha \wedge d (\beta+\delta)\Bigr),\nn
N_4 &=& {1 \over 4\pi}\int \sin [\beta-\delta] \Bigl(d \alpha \wedge d (\beta-\delta)\Bigr),
\label{eq:3.18}\ea
and express the Chern number of each band in terms of them as

\ba C_u &=& {1\over 2} ( N_1 +  N_2 ) +{1\over 4} ( N_3 + N_4 ) , \nn
C_v &=& - N_1 , \nn
C_w &=& {1\over 2} ( N_1 - N_2 ) -{1\over 4} ( N_3 + N_4 )\label{eq:3.19} .\ea
Despite its appearance we know only two of the integers can be truly independent~\cite{Simon}. In turn, this observation suggested to us a strategy to parameterize the four angles $(\alpha, \beta, \gamma, \delta)$ as mappings to two independent two-spheres in the manner described below.

\begin{figure}[t]
\includegraphics[width=80mm]{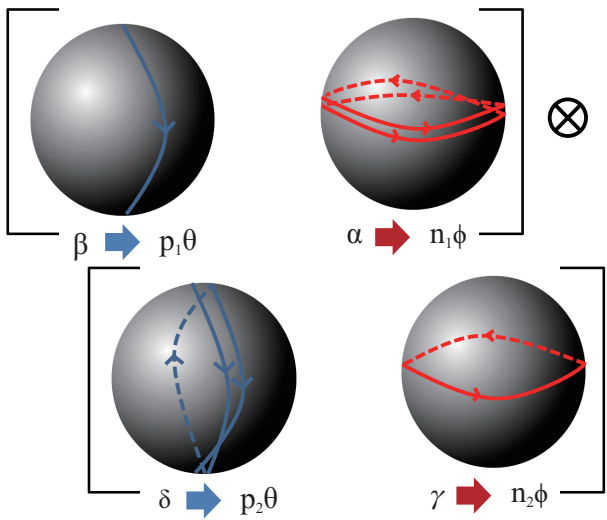}
\caption{(color online) Schematic representation of a pair of monopole charge-changing unitary operations in the three-band model. $(p_1, n_1) = (1,2)$ and $(p_2, n_2 ) = (3,1)$ in the two spheres, respectively.}
\label{fig2}
\end{figure}

Explicitly, we introduce one stereographic mapping for $(\alpha, \beta)$ by writing

\ba \hat{d}_1 = (\sin \beta \cos \alpha, \sin \beta \sin \alpha, \cos \beta) \ea
and letting $\hat{d}_1$ wrap the sphere $S^2$. Winding number of $\hat{d}_1$ is precisely the integer $N_1$ defined in Eq. (\ref{eq:3.18}). In the same way we let

\ba \hat{d}_2 = (\sin \delta \cos \gamma, \sin \delta \sin \gamma, \cos \delta) \ea
wrap another, independent two-sphere with the winding number $N_2$. Under such  prescriptions the third integer $N_3 + N_4$ is not independent, but uniquely fixed by the previous two stereographic projections. To prove this point we borrow the ideas developed from the two-band Chern number engineering and write (see Fig. \ref{fig2})

\ba (\alpha, \beta) = (n_1 \phi, p_1 \theta) , ~~~ (\gamma , \delta ) = (n_2 \phi, p_2 \theta ), \label{eq:3.25}\ea
using two pairs of integers $(p_1 , n_1)$ and $(p_2 , n_2)$. The angles $(\theta, \phi)$ span the two-sphere just once, $(1/4\pi)\int \sin \theta d\phi \wedge d\theta = 1$, and can be used to define the pristine vector $\hat{d} = (\sin \theta \cos \phi, \sin \theta \sin \phi, \cos \theta )$. According to the language developed for two-band engineering, $\hat{d}_1$ and $\hat{d}_2$ are derived from $\hat{d}$ as

\ba \hat{d}_1 = \hat{d}^{(p_1 , n_1)}, ~~ \hat{d}_2 = \hat{d}^{(p_2 , n_2 )}. \ea
With the substitution (\ref{eq:3.25}), winding numbers $N_1$ through $N_4$ are worked out as

\ba N_1 &=& {1 \over 4\pi} n_1 p_1 \int \sin p_1 \theta (d \phi \wedge d \theta) = n_1 P_1 ,\nn
N_2 &=& {1 \over 4\pi} n_2 p_2 \int \sin p_2 \theta (d \phi \wedge d \theta) = n_2 P_2 , \nn
N_3 &=& N_4 =  n_1 P_{12}  . \label{eq:3.27}\ea
Indeed $N_3$ and $N_4$ are not independent from other integers. 
The ``polarity function" $P_i = [1-(-1)^{p_i} ]/2$ and $P_{12} = [1-(-1)^{p_1 + p_2 } ]/2$ are introduced to keep track of the even ($P_i = 0$) or odd ($P_i = 1)$ polarity of the winding.  The bands' Chern numbers are related to two pairs of integers $(n_1, p_1)$ and $(n_2, p_2)$ used in the stereographic mapping as

\ba C_u & = & {1\over 2} ( n_1 P_1 + n_2 P_2 ) + {1\over 2}n_1 P_{12} , \nn
C_v &=& - n_1 P_1 , \nn
C_w &=&  {1\over 2} ( n_1 P_1 - n_2 P_2 ) - {1\over 2}n_1 P_{12}. \label{eq:3.29}\ea
One can easily convince oneself that arbitrary integers satisfying $C_u + C_v + C_w=0$ can be engineered through judicious choices of $(n_1 , p_1)$ and $(n_2 , p_2)$. Equation (\ref{eq:3.29}) is the central result of our endeavor relating the Chern numbers in the three-band model to a pair of monopole charges, or to be more precise, two pairs of integers governing two singular unitary transformations. 

With this general result we propose the following recipe to construct three-band models of arbitrary Chern numbers:

\begin{enumerate}

\item Write $(\alpha, \beta)$ and $(\gamma, \delta)$ in the unitary matrix elements according to Eq. (\ref{eq:3.25}). Other parameters can be set to zero. Choose some energy dispersion for each $e_i$. Make sure different $e_i$'s do not cross anywhere in the Brillouin zone. 

\item Choose for the pristine vector $\hat{d}$ the one with a unit winding number. One can take $\hat{d}$ from the Haldane model, for instance. 

\end{enumerate}

With the three-band Hamiltonian engineered in this way, Chern numbers of the bands can be read off from Eq. (\ref{eq:3.29}). We have tried the scheme explicitly for the pristine vector taken from the Haldane model. For all the integer combinations $(n_1, p_1)$ and $(n_2, p_2)$ we have tried, explicit computation of the Chern number for the resulting Hamiltonian over the Brillouin zone was found to be consistent with the general formulas of Eq. (\ref{eq:3.29}).

\section{Summary}

The space of unitary matrices that preserve the same set of energies $(e_1 , e_2 , e_3 )$ forms the fiber bundle $\mathrm{SU(3)}/(\mathrm{U(1)}\times\mathrm{U(1)})$. The homotopy map into this space from two dimensions reads~\cite{nakahara}

\ba && \pi_2 (\mathrm{SU(3)}/(\mathrm{U(1)}\times\mathrm{U(1)} ) \nn
&& ~~~~~ \simeq \pi_1 (\mathrm{U(1)} \times \mathrm{U(1)} ) \nn
&& ~~~~~~~~~ \simeq \pi_1 (\mathrm{U(1)} ) \times \pi_1 (\mathrm{U(1)} ) .  \label{eq:3.20}\ea
Two independent integers are thus expected from such mapping. Another relation, $\pi_2 (\mathrm{SU(2)/U(1)}) \simeq \pi_1 (\mathrm{U(1)})$, states that each of these integers is related to the Skyrmion number, or the integer of the map $S^2 \mapsto {\rm SU(2)/U(1)} \simeq S^2$, as we have just proposed. 

The goal of our work was to suggest a transparent recipe to generate arbitrary Chern numbers in a three-band model. The idea was based on the equivalence of the topological number of a given band to a certain monopole charge. Monopole charge-changing operations are none other than unitary transformations on the Hamiltonian. For the three-band case we proposed an explicit topology-engineering scheme based on the manipulation of a pair of magnetic monopole charges. Concrete realization of the unitary rotation scheme proposed here in condensed matter systems, both real and artificial~\cite{manoharan}, will be considered in the future.

\acknowledgments{J. H. H. thanks members of the MIT condensed matter theory group for their hospitality during his sabbatical leave. This work is supported by the NRF grant No. 2013R1A2A1A01006430 (JHH) and No. 2013R1A1A2058046 (GG).}

\end{document}